\documentclass[aps,onecolumn,preprintnumbers]{revtex4}
\usepackage{amsmath}
\usepackage{graphicx}
\usepackage{mathrsfs}
\usepackage{amssymb}
\usepackage{amsfonts}
\usepackage{amsmath}

\begin{document}

\title{Franck-Condon Physics in a Single Trapped Ion}
\author{Y. M. Hu$^{1,2},$ W. L. Yang$^{1},$ Y. Y. Xu$^{1,2},$ F. Zhou$^{1,2}$, L. Chen$^{1}$, K. L. Gao$^{1}$, M. Feng$^{1}$}
\email{mangfeng@wipm.ac.cn}
\author{C. Lee$^{3}$}
\email{chleecn@gmail.com}
\affiliation{$^{1}$State Key Laboratory of Magnetic Resonance and Atomic and Molecular
Physics, Wuhan Institute of Physics and Mathematics, Chinese Academy of
Sciences, and Wuhan National Laboratory for Optoelectronics, Wuhan 430071,
China}
\affiliation{$^{2}$Graduate School of the Chinese Academy of Sciences, Beijing 100049,
China}
\affiliation{$^{3}$State Key Laboratory of Optoelectronic Materials and Technologies,
School of Physics and Engineering, Sun Yat-Sen University, Guangzhou 510275,
China}

\begin{abstract}
We propose how to explore the Franck-Condon (FC) physics via a single ion
confined in a spin-dependent potential, formed by the combination of a Paul
trap and a magnetic field gradient. The correlation between electronic and
vibrational degrees of freedom, called as electron-vibron coupling, is
induced by a nonzero gradient. For a sufficiently strong electron-vibron
coupling, the FC blockade of low-lying vibronic transitions takes place. We
analyze the feasibility of observing the FC physics in a single trapped ion,
and demonstrate various potential applications of the ionic FC physics in
quantum state engineering and quantum information processing.
\end{abstract}

\maketitle

\section{Introduction}

The Franck-Condon (FC) principle is a well-known fundamental law to explain
the intensity of vibronic transitions in molecules \cite{Franck,Condon}, in
which the transition intensity is proportional to the FC factor defined by
the square of the overlap integral between the vibrational wavefunctions of
the two involved states. The FC physics actually exists in various systems
of interactions between mechanical and electronic degrees of freedom. In
particular, a very small or zero FC factor will cause transition suppression
named as the FC blockade and a nonzero FC factor between different
vibrational modes may cause vibrational sidebands~\cite{Molecule FCB,theory
FCB}. Besides the conventional experiments of molecules, the electronic
transport through quantum dots could be exponentially suppressed in the
region of strong electron-vibron coupling~\cite{QD FCB}. The conspicuous
transport property in the strong coupling regime plays an important role in
both single-molecule devices~\cite{Molecule device} and
nano-electromechanical systems~\cite{Nano system}.

To the best of our knowledge, the FC physics has never been clearly
demonstrated in a true single-particle system. In recent, an ensemble of
atoms confined within a spin-dependent optical lattice~\cite{optical lattice}
has been demonstrated for sideband cooling and coherent operations via FC
physics. That experiment can be regarded as an effective single-particle
implementation only when the inter-atom interaction can be neglected.
However, as the system is cooled to only populating the ground and
first-excited vibrational states for each lattice cell, the s-wave
scattering between atoms have to be taken into account. For a system of
shallow lattices, the atoms can easily tunnel between neighboring sites and
therefore the probability of multiple particles in a particular site will be
very significant. For a system of deep lattices, if the atomic number per
lattice site is larger than one, the on-site interaction between atoms
cannot be ignored and the single-atom model is hard to describe the system.
Even for a system of no inter-particle interaction, due to the intrinsic
nature of a spin-dependent optical lattice, the unavoidable coupling between
next-nearest-neighboring sites may destroy the FC physics and induce complex
quantum transport along the lattice axis. Moreover, due to the wavelength
and intensity limits of the optical lattices, the maximum shift and the
total number of vibronic states are both limited.

In this article, we present a proposal for observing the FC physics via a
true single-atom system, i.e., a single trapped ion, and discuss its
potential applications in quantum state manipulation. The key point is that
the electron-vibron coupling in a single trapped ion is induced and
controlled by a magnetic field gradient (MFG). Compared to the
electron-vibron coupling generated by radiation of non-resonant laser beams
on the ion, which is too weak to observe the FC physics, the MFG-induced
electron-vibron coupling is controllable and could be strong enough to
observe. We may apply this coupling to suppress or even block some undesired
transitions, called FC blockade, or to enhance some desirable transitions.
Attribute to its clear environment and high controllability, a single
trapped ion opens a new area for studying the FC physics at the single-atom
level. Beyond the fundamental interests in various fields from quantum
spectroscopy to quantum transport, the ionic FC physics is of promising
applications in quantum state engineering.

\section{Model and FC blockade}

We consider a single ultracold ion confined in a Paul trap \cite{Paul} and
the ion only populates two possible electronic levels $\left\vert \downarrow
\right\rangle $ and $\left\vert \uparrow \right\rangle $ of different
magnetic dipole moments. In usual ion trap in the absence of MFG, radiation
of laser beams on the ion with blue- and red-detuning could yield couplings
between the vibrational and electronic degrees of freedom. But this coupling
is generally weak for the ultracold ion within Lamb-Dicke limit (LDP). For
ensuring FC blockade between low-lying vibrational states, a MFG with a
sufficiently large gradient is required to enhance the electron-vibron
coupling. Without loss of generality, for a one-dimensional trap and the
gradient along the axis, the electron-vibron coupling could be expressed as
\begin{equation}
H_{0}=g\mu _{B}b\cdot \frac{\sigma _{z}}{2}\cdot \delta z=G\cdot \frac{%
\sigma _{z}}{2}\cdot (a^{\dagger }+a),
\end{equation}%
with the electron-spin $g$-factor, the Bohr magneton $\mu _{B}$ and the
phonon creation (annihilation) operators $a^{\dagger }$ $(a)$. $\delta z$ is
the oscillation amplitude of the ion along the $z$ axis. Here, $b$ denotes
the magnetic field gradient sensed by the ion, and $G=g\mu _{B}b\sqrt{\hbar
/2m\omega _{z}}$ with the ion mass $m$ and the trap frequency $\omega _{z}$.

The spin-dependent potentials can be written as%
\begin{equation}
V_{\sigma }=H_{0}+\frac{1}{2}m\omega _{z}^{2}z^{2}=\frac{1}{2}m\omega
_{z}^{2}(z+z_{0}\sigma _{z})^{2},
\end{equation}
and sketched in Fig. 1. With the electron-vibron coupling given in Eq. (1),
it is easy to obtain the shift $z_{0}=\frac{\mu_{B}b}{m\omega_{z}^{2}}$. For
a large gradient $b$, $z_{0}$ would be sufficiently large and the overlap
between the low-lying vibrational wavefunctions becomes very small. As a
result, the transition between the two vibrational ground states, $\mid
0,\downarrow\rangle\leftrightarrow \mid 0,\uparrow\rangle$, would be
strongly suppressed. This is the FC blockade.

\begin{figure}[tbh]
\begin{center}
\includegraphics[width= 0.4 \columnwidth]{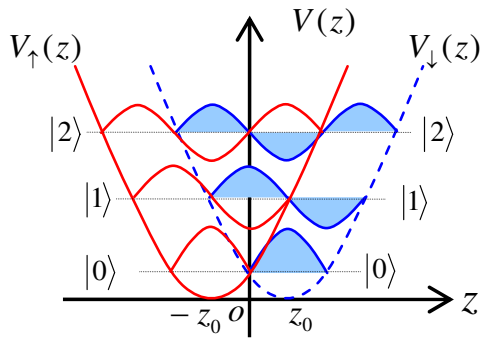}
\end{center}
\caption{Schematic diagram of the spin-dependent trap for a single ion. The
red solid and blue dashed curves are the potentials for the ion in $%
|\uparrow \rangle $ and $|\downarrow \rangle $, respectively. Here, $%
|n\rangle $ with $n=$0, 1, 2 denotes the vibrational levels. For a
sufficiently large distance between the two equilibrium positions, there is
no significant overlap between the two low-lying vibrational states so that
the FC blockade appears.}
\label{fig1}
\end{figure}

Specifically, due to the nonzero gradient, the ionic electron-vibron
coupling is governed by a new effective LDP $\eta ^{\prime
}=\sqrt{\eta ^{2}+\varepsilon ^{2}}$, where $\eta $ is the original
LDP for the case of no gradient, and $\varepsilon =\frac{\partial
\omega _{0}}{\partial z}\sqrt{\hslash /2m\omega _{z}^{3}}$ with
$\frac{\partial\omega_{0}}{\partial z}= \frac{2\mu_{B}}{\hbar}b$ is
the additional LDP caused by the gradient
\cite{Wunderlich1,Wunderlich2}. This additional LDP has been
observed in a recent experiment \cite{Wunderlich4}. Making use of
the Baker-Campbell-Hausdorff formula
$e^{A+B}=e^{A}e^{B}e^{-\frac{1}{2}[A,B]}$ (under the condition of $
[A,[A,B]]=[B,[A,B]]=0$), the transition matrix elements~\cite{theory
FCB,FC matrix1,FC matrix2,FC matrix3} between the vibrational states
$\left\vert 0,\sigma \right\rangle $ and $\left\vert n,\sigma
^{\prime }\right\rangle$ ($\sigma ,\sigma ^{\prime }=\downarrow,
\uparrow $) could be written as
\begin{align}
M_{0\rightarrow n}& =M_{\left\vert 0,\sigma \right\rangle \leftrightarrow
\left\vert n,\sigma ^{\prime }\right\rangle }=\left\langle n\right\vert
e^{i\eta ^{\prime }(a+a^{\dagger })}\left\vert 0\right\rangle ,  \notag \\
& =e^{\frac{\eta ^{\prime 2}}{2}}\left\langle n\right\vert
\sum_{j,k}\frac{(i\eta ^{\prime })^{j+k}}{j!k!}a^{j}(a^{\dagger
})^{k}\left\vert
0\right\rangle  \notag \\
& =e^{-\frac{\eta ^{\prime 2}}{2}}\frac{(i\eta ^{\prime })^{n}}{\sqrt{n!}}.
\end{align}
Thus the FC factor is $\left\vert M_{0\rightarrow n}\right\vert
^{2}=e^{-\eta ^{\prime 2}}\eta ^{\prime 2n}/n!$, which is
exponentially sensitive to the ionic electron-vibron coupling, as
plotted in Fig. 2. By adjusting the trap frequency $\omega_{z}$
and/or the gradient $b$, the effective LDP $\eta^{\prime}$ will
suppress the transition between the vibrational ground states of
different electronic levels, while allow transitions between the
vibrational ground and excited states of different electronic
levels. As shown in Fig. 2, the maximum FC factors gradually
decrease with the increase of the phonon number $n$. However, the FC
factors between different vibrational states remain non-zero, which
may cause the vibrational sidebands useful for cooling the ion.
\begin{figure}[tbh]
\begin{center}
\includegraphics[width= 0.6\columnwidth]{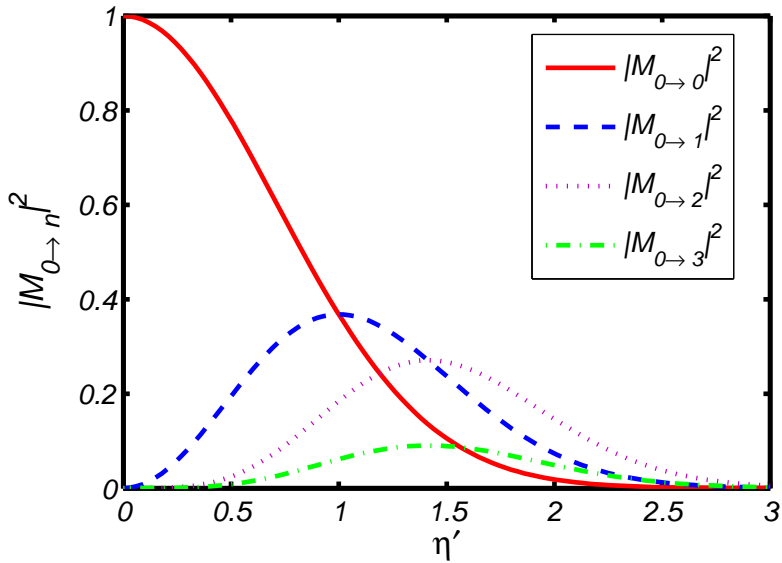}
\end{center}
\caption{The FC factors versus the new effective LDP $\protect\eta ^{\prime
} $: $\left\vert M_{0\rightarrow n}\right\vert ^{2}$ for the transitions $%
\left\vert \downarrow ,0\right\rangle \leftrightarrow \left\vert \uparrow
,n\right\rangle $ with $n=0$, $1$, $2$, $3$, respectively.}
\label{fig2}
\end{figure}

\section{Observation of FC blockade via CNOT gates}

The FC physics in a single trapped ion could be observed using a
blockade-induced controlled-NOT (CNOT) gate. To this end, we encode
the control (target) qubits in the vibrational (internal) states of
the ion. For the carrier transitions between the states $\left\vert
n,\downarrow \right\rangle$ and $\left\vert n,\uparrow \right\rangle
$, we have the Rabi
frequency%
\begin{equation}
\Omega _{n,n}=\left\vert \frac{1}{\hslash }\left\langle n,\uparrow
\right\vert H_{I}\left\vert \downarrow ,n\right\rangle \right\vert =\lambda
e^{-\frac{\eta ^{\prime 2}}{2}}|L_{n}(\eta ^{\prime 2})|,
\end{equation}%
where
\begin{equation}
H_{I}=\hbar \lambda \lbrack \sigma _{+}e^{i\eta ^{\prime }(a^{\dagger
}+a)}+\sigma _{-}e^{-i\eta ^{\prime }(a+a^{\dagger })}],
\end{equation}%
with the magnetic dipole coupling strength $\lambda $, $\sigma _{+}=\mid
\uparrow \rangle \langle \downarrow \mid $, $\sigma _{-}=\mid \downarrow
\rangle \langle \uparrow \mid $ and the Laguerre polynomial $L_{n}(x)$.

To encode two of the vibrational states as a control qubit, one may
choose a proper gradient to block the low-lying vibrational
transition and enable the high-lying vibrational transition. For
simplicity, we choose the low-lying vibrational state $\left\vert
0\right\rangle $ to be blocked and calculate the FC factors
$\left\vert M_{n\rightarrow n}\right\vert ^{2}$ for transitions
$\left\vert n,\sigma \right\rangle \leftrightarrow \left\vert
n,\sigma ^{\prime }\right\rangle $ with $n=0,1,2$, see Fig. 3 (a).
It indicates that the quantum transitions $\left\vert 0,\downarrow
\right\rangle \leftrightarrow \left\vert 0,\uparrow \right\rangle$
and $\left\vert 1,\downarrow \right\rangle \leftrightarrow
\left\vert 1,\uparrow \right\rangle $ are almost completely blocked
if $\eta ^{\prime }>3$, while the FC factor for $\left\vert
2,\downarrow \right\rangle \leftrightarrow \left\vert 2,\uparrow
\right\rangle $ is still not small up to $\eta ^{\prime }\sim 3.5$.
So in the region of $3<\eta ^{\prime }\leq 3.5$, we may encode the
vibrational states $|0\rangle $ and $|2\rangle $ as the control
qubit. By driving the carrier transition for a duration $\tau $ satisfying $%
\Omega _{2,2}\tau =\pi /2$, the states $\mid \downarrow \rangle $ and $\mid
\uparrow \rangle $ are interchanged only if the vibrational state is $%
|2\rangle $. Then we could obtain that $\Omega _{0,0}/\Omega _{2,2}=1/\left(
1-2\eta ^{\prime 2}+\eta ^{\prime 4}/2\right) <0.0425$ if $\eta ^{\prime }>3$
. This means that, comparing to $\Omega _{2,2}\tau $, $\Omega _{0,0}\tau $
can be ignored. Therefore, we may construct a CNOT gate in the subspace
spanned by $\left\vert 0,\downarrow \right\rangle $, $\left\vert 0,\uparrow
\right\rangle $, $\left\vert 2,\downarrow \right\rangle $ and $\left\vert
2,\uparrow \right\rangle $,
\begin{equation}
U=\left(
\begin{array}{cccc}
\cos \Omega _{0,0}\tau & i\sin \Omega _{0,0}\tau & 0 & 0 \\
i\sin \Omega _{0,0}\tau & \cos \Omega _{0,0}\tau & 0 & 0 \\
0 & 0 & \cos \Omega _{2,2}\tau & i\sin \Omega _{2,2}\tau \\
0 & 0 & i\sin \Omega _{2,2}\tau & \cos \Omega _{2,2}\tau%
\end{array}%
\right) .
\end{equation}%
Eliminating the undesired phase factors on $|2\rangle $, we obtain a CNOT
gate on the electron states conditional on the vibrational states of the
ion: $|0,\downarrow (\uparrow )\rangle \rightarrow |0,\downarrow (\uparrow
)\rangle $ and $|2,\downarrow (\uparrow )\rangle \rightarrow |2,\uparrow
(\downarrow )\rangle $. The fidelity of the CNOT gate is estimated with
respect to the effective LDP $\eta ^{\prime }$ in Fig. 3(b), in which the
growth of $\eta ^{\prime }$ improves the fidelity.

\begin{figure}[tbh]
\begin{center}
\includegraphics[width= 1.0\columnwidth]{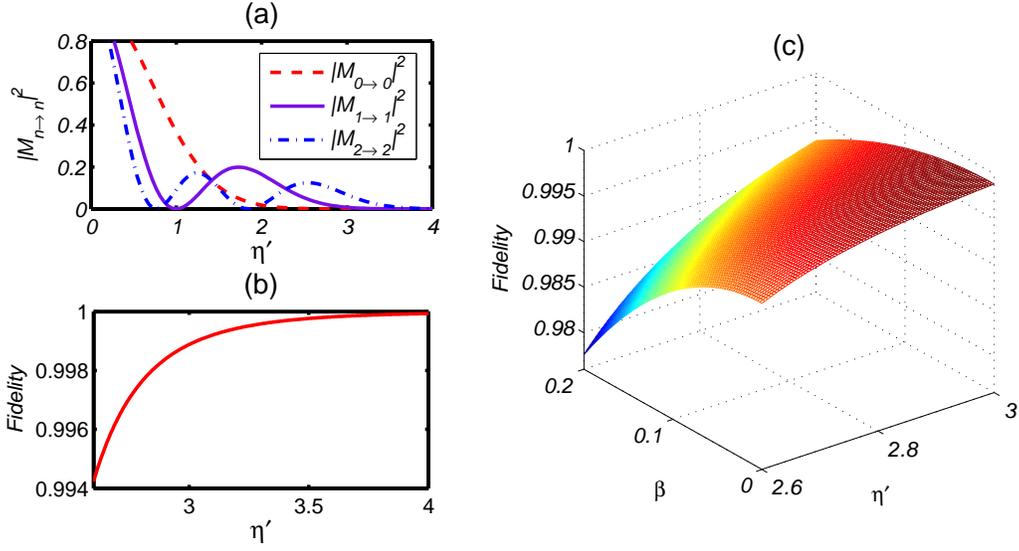}
\end{center}
\caption{(a) FC factors $\left\vert M_{n\rightarrow n}\right\vert ^{2}$ for
transitions $\left\vert n,\protect\sigma \right\rangle \leftrightarrow
\left\vert n,\protect\sigma ^{\prime }\right\rangle $ versus $\protect\eta %
^{\prime }$. (b) Fidelity of the CNOT gating versus $\protect\eta %
^{\prime }$ for the initial vibrational state $\left( |0\rangle +|2\rangle
\right) /\protect\sqrt{2}$. (c) Fidelity of performing the CNOT gate versus $%
\protect\eta ^{\prime }$ and the error factor $\protect\beta $, where the
initial vibrational state is $\left[ \protect\alpha \left( |0\rangle
+|2\rangle \right) +\protect\beta |1\rangle \right] /\protect\sqrt{2}$.}
\label{fig3}
\end{figure}

The FC physics is also observable in the mediate region of $2.6\leq \eta
^{\prime }<3$, where the FC factor for $|0,\downarrow \rangle
\leftrightarrow |0,\uparrow \rangle $ is almost zero, but the FC factors for
$|1,\downarrow \rangle \leftrightarrow |1,\uparrow \rangle $ and $%
|2,\downarrow \rangle \leftrightarrow |2,\uparrow \rangle $ are still
significant. Because $\left\vert M_{2\rightarrow 2}\right\vert
^{2}>\left\vert M_{1\rightarrow 1}\right\vert ^{2}$, we may still encode the
control qubit in $|0\rangle $ and $|2\rangle $. The CNOT gate could still be
expressed by Eq. (6) and its fidelity versus $\eta^{\prime}$ is shown in
Fig. 3(b). Compared to the case of larger gradients, an unfavorable effect
may appear as a result of the unwanted population on the vibrational state $%
|1\rangle$. A simple estimation of the detrimental influence from the
undesired population on $|1\rangle $ is obtained in Fig. 3(c) by assuming
the initial vibrational state as $\left[\alpha\left( |0\rangle
+|2\rangle\right) +\beta |1\rangle\right] /\sqrt{2}$ with the error factor $%
\beta$.

When the gradient is tuned into the region $0<\eta^{\prime }\le 1$, the FC
blockade does not happen for the ground vibrational transition but for other
high-lying vibrational transitions at some special points, such as the
blockade of $\left|1,\downarrow\right\rangle \leftrightarrow
\left|1,\uparrow\right\rangle$ at $\eta^{\prime}=1$ and the blockade of $%
\left|2,\downarrow\right\rangle \leftrightarrow
\left|2,\uparrow\right\rangle $ at $\eta^{\prime}=0.765$. This reminds us of
the 'magic' LDP mentioned in \cite{Monroe}. Therefore, tuning the effective
LDP to the `magic' point $\eta^{\prime}=1$, we may employ the vibrational
states $\left|0\right\rangle$ and $\left|1\right\rangle$ to encode the
control qubit. The CNOT gate is then implemented in the subspace spanned by $%
\left|0,\downarrow\right\rangle$, $\left|0,\uparrow\right\rangle$, $%
\left|1,\downarrow\right\rangle$ and $\left|1,\uparrow\right\rangle$, where
the internal states flip only when the vibrational state is $%
\left|0\right\rangle$. Alternatively, tuning the effective LDP to the other
`magic' point $\eta^{\prime}=0.765$, we may encode the control qubit in the
vibrational states $\left|0\right\rangle$ and $\left|2\right\rangle$.
Different from the schemes for large and mediate gradients, our CNOT gate in
the region of small gradients can only be accomplished at some 'magic'
points. As a result, the quality of the performance is very sensitive to the
effective LDP. We have estimated this sensitivity and the relations of the
MFG to the LDP in Fig. 4.

\begin{figure}[tbh]
\begin{center}
\includegraphics[width= 0.75\columnwidth]{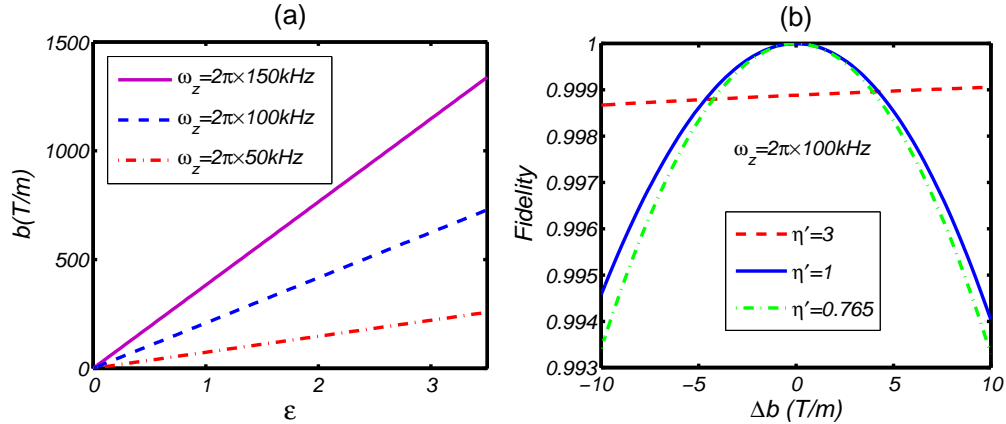}
\end{center}
\caption{(a) The required gradient $b$ versus the additional LDP $\protect%
\varepsilon $ for different trap frequencies. (b) Fidelity of
CNOT gating versus the fluctuation of gradient $\Delta b$ in the case of $%
\protect\omega _{z}=2\protect\pi \times 100$ kHz: red dashed curve for $%
b=624 $ T/m (i.e. $\protect\eta ^{\prime }=3$); blue solid curve for $b=208$
T/m (i.e. $\protect\eta ^{\prime }=1$) and the green dashed-dotted for $%
b=159.2$ T/m (i.e. $\protect\eta ^{\prime }=$0.765).}
\label{Fig4}
\end{figure}

In our CNOT proposal, since two motional states of the ion are encoded as
the control qubit, it requires ground-state cooling, which could be made
before the MFG is applied or in the presence of a large MFG. As we will
briefly discuss later, an appropriately big MFG sometimes helps for laser
cooling.

\section{Applications in quantum state engineering}

\subsection{Preparation of Fock states}

The observable FC physics could be used to prepare motional Fock states in a
probabilistic way. We consider the initial state%
\begin{equation}
\left\vert \Psi (0)\right\rangle =\left\vert \downarrow \right\rangle
\left\langle \downarrow \right\vert \left( P_{0}\left\vert 0\right\rangle
\left\langle 0\right\vert +P_{1}\left\vert 1\right\rangle \left\langle
1\right\vert +P_{2}|2\rangle \left\langle 2\right\vert \right) ,
\end{equation}%
with the population probability $P_{k}$ and the average phonon number $%
P_{1}+2P_{2}<1$. To prepare the ground motional state $|0\rangle $, one has
to eliminate the populations in states $|1\rangle $ and $|2\rangle $. By
tuning the gradient to block the transition $\left\vert 1,\downarrow
\right\rangle \leftrightarrow \left\vert 1,\uparrow \right\rangle $, the
population in $|1\rangle $ could be screened away via a $\pi /2$ carrier
transition pulse following a measurement on $\left\vert \uparrow
\right\rangle $. Similarly, by tuning the gradient to block the transition $%
\left\vert 2,\downarrow \right\rangle \leftrightarrow \left\vert 2,\uparrow
\right\rangle $, the population in $|2\rangle $ could be screened away via a
$\pi /2$ carrier transition pulse following a measurement on $\left\vert
\downarrow \right\rangle $. In above operations, the generation of the
expected Fock state is probabilistic. So we have to employ the
repeat-until-success method \cite{RE}. Once the desired internal state is
successfully detected, the desired motional state is prepared with a unity
fidelity.

\subsection{Modification in single-qubit gate operations}

Due to the existence of the FC blockade in the regime of large MFG,
some new difficulties for single-qubit operations in a string of
trapped ions appear. The proposals \cite{Wunderlich1, Wunderlich2}
from Wunderlich's group focused on the regime of small MFG, which
corresponds to a small LDP. Under such a small LDP, the first-order
expansion works very good and the ground-state cooling is indeed not
necessary. In the regime of higher MFG supporting the FC blockade,
the large MFG favors a working Ising coupling within a shorter time,
which makes the two-qubit conditional operations faster. However,
since the large MFG strongly suppresses some vibrational
transitions, it becomes more difficult to perform the Hadamard gates
via carrier transitions.

We show a specific simulation for a spin flip in the presence and absence of
MFG, see Fig. 5. In our simulation, the single trapped ion within the
initial internal state $|\downarrow \rangle$ and thermal motional state
(e.g., $\langle n\rangle =5$ or 0.1) under a MFG ($\eta ^{\prime }=1$)
evolves according to the dynamical population
\begin{equation}
P_{\downarrow }(t)=\left\langle \downarrow \right\vert
Tr_{n}[e^{-iH_{I}t/\hbar }\rho (0)|\downarrow \rangle \left\langle
\downarrow \right\vert e^{iH_{I}t/\hbar }]\left\vert \downarrow
\right\rangle ,
\end{equation}
with $\rho (0)=\sum\limits_{n=0}M_{n}|n\rangle \langle n|$, $M_{n}=N[\langle
n\rangle /(1+\langle n\rangle )]^{n}$ and N the normalization constant.
\begin{figure}[tbh]
\begin{center}
\includegraphics[width= 0.5\columnwidth]{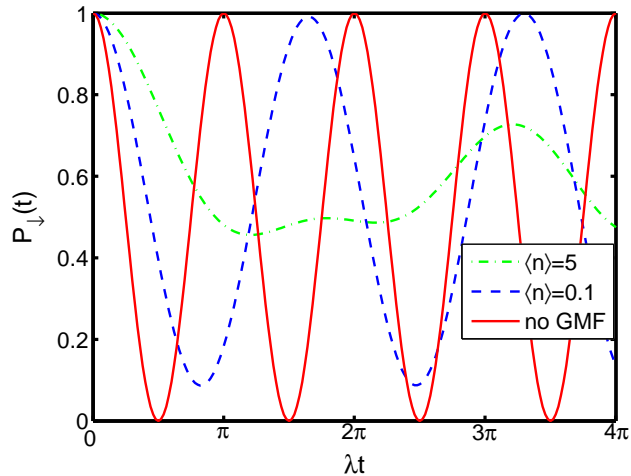}
\end{center}
\caption{Time evolution of the population of the ion's internal states $%
|\downarrow \rangle $ under spin-flip gate in the absence and presence of
MFG, where we set $\langle n\rangle =5$ or 0.1, and $\protect\eta ^{\prime
}=1$.}
\label{Fig5}
\end{figure}
Fig. 5 indicates the necessity to cool the ion to the vibrational ground
state in the presence of a large MFG. A possible solution is to increase the
Rabi frequency by enhancing the radiation. Since the suppression is
sensitive to $\eta^{\prime}$, the appropriate Rabi frequency could be
determined by interrogative pulse operations with respect to different
values of $\eta^{\prime}$. This result is also applicable to the refocusing
pulses \cite{refocusing1} for removing the undesired couplings due to the
Ising coupling \cite{refocusing2}. As the refocusing pulses are based on
carrier transitions, like the Hadamard gate above, stronger pulses are
necessary for refocusing under a large MFG. Alternatively, we may switch off
the MFG when performing single-qubit gates. To this end, employment of
lasers, instead of microwaves as in \cite{Wunderlich1, Wunderlich2}, is
necessary for individually accomplishing the single-qubit operations, where
the FC physics does not work in the carrier transition and cooling to the
vibrational ground state is unnecessary.

\section{Experimental feasibility and challenge}

For a real experimental implementation of the CNOT gate, we may employ the
hyperfine levels $\left\vert S_{1/2},F=0,m_{F}=0\right\rangle $ and $%
\left\vert S_{1/2},F=1,m_{F}=1\right\rangle $ of $^{171}$Yb$^{+}$ as $%
\left\vert \downarrow \right\rangle $ and $\left\vert \uparrow \right\rangle
$, respectively. Here, the transition frequency for $\left\vert \downarrow
\right\rangle \leftrightarrow \left\vert \uparrow \right\rangle $ is $\omega
_{0}\approx 2\pi \times 12.6$ GHz \cite{Wunderlich1,new1}. With the $z$-axis
trap frequency $\omega _{z}=2\pi \times $100 kHz and magnetic dipole
coupling strength $\lambda =2\pi \times 50$ kHz, the CNOT gate will be
accomplished within 19.2 $\mu $s, 8.2 $\mu $s, and 6.7 $\mu $s for $\eta
^{\prime }=$3, 1 and 0.765, respectively. To perform the CNOT gates highly
coherently, we require the coherence time of the employed hyperfine levels
to be at least longer than 192 $\mu $s. Fortunately, the latest experiment
has shown that this coherence time could be 5 ms \cite{explain}.

To achieve our scheme, moreover, we may employ the weak region of $\eta
^{\prime }$, e.g., $\eta ^{\prime }=1$. To this end, due to sensitivity to
the `magic' points, a highly stable gradient is required for achieving the
high-fidelity operations. We have estimated the sensitivity to the
fluctuation of the gradient, see Fig. 4(b) where the fidelities are larger
than 0.994 for all schemes if the gradient fluctuation $|\Delta b|\leq 10$
T/m. Compared to \cite{Monroe} with the `magic' LDP controlled by the
wave-vector, phase and intensity of the radiating lasers, the CNOT operation
in our scheme is mainly governed by the MFG, whose strength and stability
are key to the implementation.

With currently available technologies, a big challenge of our scheme is how
to realize a large MFG. Current-carrying coils in an anti-Helmholtz-type
arrangement \cite{Wunderlich1, new5} and permanent magnets \cite{Wunderlich4}
have been applied to attain a gradient up to tens of T/m. To achieve a
working Ising type interaction between two ions separated by few
micrometers, the MFG is required to be on the order of 100 T/m \cite{feng,
new-wun}. Although this is still challenging with current techniques, there
are some efforts toward this aim using new materials and improved designs
\cite{wang}.

\section{Conclusions}

In summary, we have explored the FC physics in a single trapped ion under an
external MFG and discussed the experimental feasibility. Although our
discussion above focused on the FC blockade, the strong electron-vibron
coupling would probably be useful for quantum simulation of, e.g., Dirac
equation \cite{dirac} or quantum walk \cite{blatt}. We argue that our study
would be useful for further understanding FC physics and its application.

Moreover, the FC physics with strong electron-vibron coupling might also be
useful for sideband cooling of the trapped ion, in which vibrational
sidebands play the important role and the carrier transitions are excluded.
In usual schemes, the cooling does not happen if the cooling laser is
orthogonal to the trapping direction \cite{EPL}. However, for an ion in a
spin-dependent potential as shown in Fig. 1, the electron-vibron coupling is
caused by the MFG, instead of the cooling laser itself. As a result, the
cooling could work even for a red-detuning beam perpendicular to the
trapping direction. However, due to FC blockade, the cooling down to the
ground motional state is sometimes impossible, but deterministically to some
certain motional states, e.g., to $n=1$ for $b=208$ T/m with $\nu =$100 kHz.
Nevertheless, using the conventional laser cooling techniques plus a MFG in
parallel, the cooling efficiency should be enhanced. In recent, there
appears a work on enhanced cooling via MFG~\cite{cooling}, in which the FC
physics has not been specifically mentioned.

\section*{Acknowledgements}

YMH is grateful to Zhangqi Yin and Qiong Chen for their helps. This work is
supported by the 100-Talent Program of Sun Yat-Sen University and the
National Natural Science Foundation of China under Grants No. 10974225, No.
11004226 and No. 11075223.


\begin{thebibliography}{99}
\bibitem{Franck} Franck J 1926 \textit{Trans. Farad. Soc.} \textbf{21} 536

\bibitem{Condon} Condon E 1926 \textit{Phys. Rev.} \textbf{28} 1182

\bibitem{Molecule FCB} Koch J and von Oppen F 2005 \textit{Phys. Rev. Lett.}
\textbf{94} 206804

\bibitem{theory FCB} Koch J, von Oppen F and Andreev A V 2006 \textit{Phys.
Rev. B} \textbf{74} 205438

\bibitem{QD FCB} Leturcq R, Stampfer C, Inderbitzin K, Durrer L, Hierold C,
Mariani E, Schultz M G, von Oppen F and Ensslin K 2009 \textit{Nat. Phys.}
\textbf{5} 327

\bibitem{Molecule device} Joachim C, Gimzewski J K and Aviram A 2000 \textit{Nature (London)}
\textbf{408} 541

\bibitem{Nano system} Ekinci K L, and Roukes M L 2005 \textit{Rev. Sci.
Instrum.} \textbf{76} 061101

\bibitem{optical lattice} Fr\"{o}rster L, Karski M, Choi J -M, Steffen A, Alt W, Meschede D,
Widera A, Montano E, Lee J H, Rakreungdet W and Jessen P S  2009 \textit{Phys. Rev. Lett.}
\textbf{103}, 233001

\bibitem{Paul} Paul W 1990 \textit{Rev. Mod. Phys.} \textbf{62} 531

Ghosh P K 1995 \textit{Ion traps} (Oxford: Clarendon Press)

\bibitem{Wunderlich1} Mintert F and Wunderlich C 2001 \textit{Phys. Rev.
Lett.} \textbf{87} 257904

\bibitem{Wunderlich2} Wunderlich C, Balzer C, Hannemann T, Mintert F,
Neuhauser W, Rei\ss\ D\ and Toschek P E 2003 \textit{J. Phys. B: At. Mol.
Opt. Phys.} \textbf{36} 1063

\bibitem{Wunderlich4} Johanning M, Braun A, Timoney N, Elman V, Neuhauser W
and Wunderlich C 2009 \textit{Phys. Rev. Lett.} \textbf{102} 073004

\bibitem{FC matrix1} Aji V, Moore J E and Varma C M 2003 arXiv:0302222

\bibitem{FC matrix2} Koch J, von Oppen F, Oreg Y and Sela E 2004 \textit{\
Phys. Rev.} B \textbf{70} 195107

\bibitem{FC matrix3} Koch J and von Oppen F 2005 \textit{Phys. Rev. }B
\textbf{72} 113308

\bibitem{Monroe} Monroe C, Leibfried D, King B E, Meekhof D M, Itano W M,
and Wineland D J 1997 \textit{Phys. Rev. }A \textbf{55} R2489

\bibitem{RE} Lim Y L, Barrett S D, Beige A, Kok P and Kwek L C 2006 \textit{%
\ Phys. Rev}. A \textbf{73} 012304

\bibitem{refocusing1} Nielsen M A and Chuang I L 2000 Quantum Computation
and Quantum Information (Cambridge: Cambridge University Press)

Gruska J 1999 Quantum Computing (Maidenhead: McGraw-Hill)

\bibitem{refocusing2} Mc Hugh D and Twamley J 2005 \textit{Phys. Rev.} A
\textbf{71} 012315

\bibitem{new1} Balzer C, Braun A, Hannemann T, Paape C, Ettler M, Neuhauser
W, and Wunderlich C 2006 \textit{Phys. Rev.} A \textbf{73} 041407(R)

\bibitem{explain} Private communication with ion trap group at Siegen
university.

\bibitem{new5} Wunderlich C and Balzer C 2003 \textit{Adv. At. Mol. Opt.
Phys.} \textbf{49} 293

\bibitem{feng} Feng M, Xu Y Y, Zhou F and Suter D 2009 \textit{Phys. Rev.} A
\textbf{79} 052331

\bibitem{new-wun} Wunderlich H, Wunderlich C, Singer K, and Schmidt-Kaler F
2009 \textit{Phys. Rev.} A \textbf{79} 052324

\bibitem{wang} Wang K, Johanning M, Feng M, Mintert F, Wunderlich C 2010
arXiv:1101.0404

\bibitem{dirac} Gerritsma R, Kirchmair G, Z\"{a}hringer F, Solano E, Blatt R
and Roos C F 2010 \textit{Nature(London)} \textbf{463} 68

\bibitem{blatt} Z\"{a}hringer F, Kirchmair G, Gerritsma R, Solano E, Blatt R
and Roos C F 2010 Phys. Rev. Lett. \textbf{104} 100503

\bibitem{EPL} Morigi G, Cirac J I, Lewenstein M and Zoller P 1997 \textit{\
Europhys. Lett.} \textbf{39} 13

\bibitem{cooling} Albrecht A, Retzker A, Wunderlich C and Plenio M B 2010
arXiv:1009.2441
\end{thebibliography}
\end{document}